\documentstyle[12pt,psfig,a4]{article}
\pssilent
\begin{document}
\begin{center}
{\large{The Role of trapped neutrino in dense stellar matter and kaon 
condensation}} \\
\vskip 1.5cm
Ruma Dutta \\
Department of Physics, Indian Institute of Technology\\
Powai, Mumbai 400 076 \\
e-mail:ruma$@$phy.iitb.ernet.in \\
\end{center}
\vskip 1.0cm
\leftline\bf{Abstract:}\\
  We investigate the effect of neutrino trapping on the kaon condensation 
process and Equation of State (EOS) in a newly formed neutron star matter 
which is less than several seconds old.   
 Using nonlinear relativistic mean field model
,we find that the presence of neutrino shifts the threshold density for kaon
 condensation and muon production to much higher density.  
We also studied the pressure and energy density of the system and found that 
the presence of neutrino stiffens the EOS 
which may be responsible for the delayed exploision mechanism for supernovae. 
\vskip 2.0cm
    
{\bf I. Introduction:}\\
\vskip\baselineskip
\noindent
  The presence of neutrino is an essential part of gravitational collapse,
  supernovae and protoneutron stars. The scenario of type II supernovae 
  is strongly marked by the dynamical behaviour of stellar medium under 
 neutrino trapping effects. During the supernovae collapse phase, a large 
 number of neutrinos are being produced by the electron capture process. So, 
 just after few seconds of a new born protoneutron star, the neutrino mean 
free path becomes larger than collapsing time scale marked by $\tau_{\rm h}
= {(G\rho)}^{\rm {1/2}}$   [ G is the gravitational const, $\rho$ $\to$ density
of the core ] . Beyond densities of about $10^{\rm {13}}$ g.$cm^{\rm {-3}}$,
 neutrinos being trapped within the matter are unable to propagate on dynamical
scale. These trapped neutrinos
become an important ingredient in discussing the stellar matter Equation of 
State (EOS). So, in this case the beta equilibrium condition is being altered  i.e \\
      \[  n + \nu_{\rm e}  \rightleftharpoons   p + e  \]    is established in
 a characteristic
of time  scale  $\tau_{\rm \beta}$ = < $\sigma_{\rm 0}c{\rho_{\rm \nu}}$ =
$10^{\rm {-15}} - 10^{\rm {-18}}$ s,  which is a function of trapped neutrino
density $\rho_{\rm {\nu}}$ ( $\sigma_{\rm 0}$ $\to$ typical cross section for $\nu$-n
process). Thus the composition of matter is affected and the threshold 
density for kaon condensation gets affected. \\
\noindent  The inner core continues to collapse and higher values of nuclear
densities are reached. A shock front is generated at the reversion of inner
core collapse. So, characteristic equation of state determines the shock 
front. In the presence of neutrinos, the threshold density for occuring 
kaon condensation is shifted to higher density. So, the role of neutrino 
trapping is claimed to explain the delayed exploision mechanism. Properties of 
trapped neutrinos are already been discussed in many works [1,2.]. \\
\parindent 3pt Here we discuss the kaon condensation and other properties of neutron star matter under the trapped neutrino condition. \\
\parindent 3pt The idea that above some critical density, the ground 
state of baryonic matter may contain  a Bose-Einstein condensate of negatively
charged kaons was given first by Kaplan and Nelson [3]. Using an SU(3)XSU(3)
chiral lagrangian , they showed that around $\rho$ $\simeq$  3$\rho_{\rm 0}$
 [  $\rho_{\rm 0}$ $\to$  equilibrium nuclear matter density ], kaon condensation
is energetically favourable. Physically, the attraction between $K^{\rm -}$
mesons and nucleons increases the density and lowers the energy of the zero
 momentum state . A condensate forms when this energy becomes equal to the kaon
chemical potential $\mu$.  \\
\parindent 3pt The presence of kaon condensation softens the equation of state 
which lowers the maximum mass of neutron star. In neutrino free dense matter,
the density at which this condensation takes place is typically 
$\rightleftharpoons$
 4$\rho_{\rm 0}$ ( $\rho_{\rm 0}$ denotes the equilibrium density)
which is much less than the density of the core of neutron star. So, 
$K^{\rm -}$ condensate is expected to be present in the core of the star. \\
Brown and Kubodera group [4] attempted the kaon condensation through s-wave 
interaction of
kaons with nucleons and found rapid rise of condensate amplitude with
density. Later few attemts [5,6]
 have been made to investigate the possibility of kaon condensation 
in the core of neutron star through p-wave interaction of kaons with nucleons
and found the small values of kaon amplitude.
Recently, Prakash group  [7] improved this analysis of kaon 
condensation of
neutron star matter including hyperons from
relativistic mean field model. They also observed the rapid rise of condensate
amplitude beyond threshold and the effect of hyperons on the threshold which 
shifts the threshold for condensation towards higher density. Very recently,
Chiapparini et al [8] studied the EOS of neutron star matter under trapped neutrino
without kaon condensation and hyperons. They studied the matter with different
electron-lepton fraction value and found almost no effect on Equation of State.
\\
\parindent 3pt Here in this work, we extend our analysis of neutron star matter
(n-p-e-$\mu$) including kaon condensation under neutrino 
trapping condition. We study the effect of neutrino trapping on the threshold 
density for kaon condensation in the core of dense stellar matter. \\
\parindent 3pt We study the matter from relativistic mean field model where
    the hadronic and leptonic densities are calculated in a self consistent 
way in $\beta$ equilibrium under two conditions. One is, charge neutrality
and the second one being the total baryon number conservation. \\
\vskip 0.2in
\noindent
{\bf II. The lagrangian } \\
\vskip\baselineskip
\noindent The hadronic fields are considered here in relativistic mean field
 model in which baryons interact via the exchange of $\sigma$-, $\omega$- 
and $\rho$- mesons with nonlinear self energy terms. \\ 
The total lagrangian is, \\
\[ \cal {L_{\rm total}} = \cal {L_{\rm B}} + \cal {L_{\rm M}} 
                          + \cal {L_{\rm L}}              \]
where $\cal {L_{\rm B}}$, $\cal {L_{\rm M}}$ and $\cal {L_{\rm L}}$ describe
the baryonic, mesonic and leptonic sector respectively and given by, \\
\[  {\cal {L_{\rm B}}} = \sum_{\rm B}{\overline{\psi_{\rm B}}}
(i \gamma^{\rm {\mu}}{\partial_{\rm {\mu}}} -g_{{\omega}B}
\gamma^{\rm {\mu}}{\omega_{\rm {\mu}}} - g_{\rm {\rho}B}\gamma^{\rm \mu}b_{\rm
 \mu}.{\tau} - m_{\rm B} + g_{\rm {\sigma}B}{\sigma}){\psi_{\rm B}}  \]
$m_{\rm B}$ $\to$ vacuum baryon mass. \\
In the mesonic sector, \\
\begin{eqnarray}
 {\cal {L_{\rm M}}} &=& {1 \over 2}(\partial_{\rm \mu}\partial^{\rm \mu}\sigma
 - {m_{\rm \sigma}}^{\rm 2}{\sigma^{\rm 2}}) -{1 \over 3}bM({g_{\rm \sigma}
{\sigma}}^{\rm 3}) -{1 \over 4}c({g_{\rm \sigma}\sigma}^{\rm 4}) \nonumber \\
&-& {1 \over 4} F_{\rm {\mu\nu}}F^{\rm {\mu\nu}} + {1 \over 2}{m_{\rm \omega}}^
{\rm 2}{\omega_{\rm \mu}}{\omega^{\rm \mu}} - {1 \over 4}B_{\rm {\mu\nu}}
B^{\rm {\mu\nu}} \nonumber \nonumber \\
 &+& {1 \over 2}{m_{\rm \rho}}^{\rm 2}{\rho_{\rm \mu}}
{\rho^{\rm \mu}}  + {\cal{L_{\rm K}}} \nonumber  
\end{eqnarray}

\[   {\cal {L_{\rm L}}} = \sum_{\rm {l=e,\mu, \nu}}{\overline{\psi_{\rm l}}} \]
We treat the meson in the mean field approximaion  and kaplan-Nelson 
SU(3)XSU(3) chiral lagrangian and kaon-baryon interaction for kaon. \\
\begin{eqnarray}
 {\cal {L_{\rm K}}} &=& {1 \over 4}f^{\rm 2}Tr{\delta_{\rm \mu}}\delta^{\rm 
\mu}cTr m_{\rm q}(u+u^{\rm \dag}-2) + iTr{\overline{\psi_{\rm B}}} 
{\gamma^{\rm \mu}}[V_{\rm \mu}, {\psi_{\rm B}}] \nonumber \\
&+& a_{\rm 1}Tr \overline{B}( {\xi}m_{\rm q}{\xi} + h.c)B	
+ a_{\rm 2}tr{\overline{B}}B({\xi}m_{\rm q}{\xi} + h.c)
+ a_{\rm 3} \lbrace Tr({\overline B}B) \rbrace	\nonumber
\end{eqnarray}
U is the nonlinear field involving pseudoscalar meson octet which retains
only $K^{\rm +}$ contributions. \\
\parindent 3pt f $\to$ is the pion decay const and C, $a_{\rm 1}$, $a_{\rm 2}$
, $a_{\rm 3}$ are constants. \\ 
  We have taken $a_{\rm 1}m_{\rm s}$ = -67.0 MeV; $a_{\rm 2}m{\rm s}$ =134.0
MeV and $a_{\rm 3}m_{\rm s}$ = - 222.0 MeV. \\
The effective mass ${m_{\rm B}}^{\rm *}$ in the presence of kaon 
condensate is given by, \\
 
  \[ {m_{\rm B}}^{\rm *} = m_{\rm B} + g_{\rm {\sigma}B}{\sigma}
 + [(a_{\rm 1}+a_{\rm 2})( 1 + Y_{\rm B}q_{\rm B}) + (a_{\rm 1} 
  - a_{\rm 2})(q_{\rm B} - Y_{\rm B}) + 4a_{\rm 3}] m_{\rm s}sin^{\rm 2}
{1 \over 2}{\theta}  \]
and the chemical potential $\mu_{\rm B}$  is given by, \\
\[ \mu_{\rm B} = \nu_{\rm B} + g_{\rm {\omega}B}{\omega}_{\rm 0}
 + g_{\rm {\rho}B}{\tau}_{\rm {3B}}b_{\rm 0} - ( Y_{\rm B} + q_{\rm B})
{\mu}sin^{\rm 2}{1 \over 2}\theta    \]
where $\tau_{\rm {3B}}$ is the $\cal z$-component of isospin of the baryon
( here only p and n) and $\nu_{\rm B}$ is given in terms of fermi momentum
 $k_{\rm {FB}}$
through \\
  \[  \nu_{\rm B} = \sqrt{ {k_{\rm {fB}}}^{\rm 2} + 
{{m_{\rm B}}^{\rm *}}^{\rm 2}} \]
  Here $m_{\rm s}$ is the scalar meson mass and $\theta$ is the kaon condensate amplitude. \\
 Thus the particle  density $\rho_{\rm B}$ are obtained as, \\
  \[ \rho_{\rm B} = {\gamma \over {6\pi^{\rm 2}}} {k_{\rm F}^{\rm 3}} \]
The degeneracy factor $\gamma$ = 2 for massive particle and =1 for neutrino. \\
and for kaon the density is,  \\
\[ \rho_{\rm k} = f^{\rm 2}({\mu}sin^{\rm 2}\theta + 4esin^{\rm 2}{1 \over 2}
\theta )   \]
with 
 \[   e = \sum_{\rm B} ( Y_{\rm B}
+ q_{\rm B}){\rho_{\rm B}}/(4f^{\rm 2})   \]
For meson, \\
\[ \mu_{\rm l}=\sqrt{{k_{\rm l}}^{\rm 2} + {m_{\rm l}}^{\rm 2}} \] l = e,$\mu$\\ 
and  $\mu_{\rm {\nu_{\rm e}, \bar{\nu_{\rm \mu}}}}$ = $k_{\rm {\nu_{\rm e},
\bar{\nu_{\rm \mu}}}}$  
\vskip 1.0cm
Charge neutrality for the matter is conserved  , \\
\[ \rho_{\rm p} = \rho_{\rm e} + \rho_{\rm {\mu}} + \rho_{\rm k} \]
\parindent 2pt we explore the effect of neutrino trapping by fixing 
electron-lepton fraction , \\
\[  Y_{\rm {le}} = { \rho_{\rm e} + \rho_{\rm {\nu_{\rm e}}} \over 
 \rho_{\rm B} } = Y_{\rm e} + Y_{\rm {\nu_{\rm e}}}                   \]
at total baryon density. \\
parindent 2pt So, we solve the $\cal E$uler-$\cal L$agrange equation for the 
baryon and meson field where the scalar density is obtained from, \\
\[  {\rho_{\rm B}}^{\rm s} = { \gamma \over 2{\pi}^{\rm 2} }
\int {{m_{\rm B}}^{\rm *} \over \sqrt{{k_{\rm B}}^{\rm 2} + {m_{\rm B}}^{\rm 2}
}} k^{\rm 2} dk          \]
the condensate amplitude $\theta$ is found from , \\
\[ \mu^{\rm 2}Cos{\theta} + 2e{\mu} -{m_{\rm k}}^{\rm 2} - d_{\rm 1} = 0 \]
where  \\
\[ 2f^{\rm 2}d_{\rm 1} = \sum_{\rm {B = n,p}} [(a_{\rm 1} + a_{\rm 2}(
1 + Y_{\rm B}q_{\rm B}) + (a_{\rm 1} -a_{\rm 2})(q_{\rm B} - Y_{\rm B})
+4a_{\rm 3}]m_{\rm s} {\rho_{\rm B}}^{\rm s}   \] 
Here we have used the coupling parameter values taken from Ref.[9].\\
For leptons, the elctron and muon are governed by the chemical potential 
$\mu_{\rm e}$ = $\mu_{\rm {\mu}}$ = $\mu$. \\
The threshold density $\rho_{\rm c}$ for condensation is determined by
setting $\theta$ = 0 in the above equation, i.e \\
\[ \mu^{\rm 2} + 2e{\mu} -{m_{\rm k}}^{\rm 2} -[2a_{\rm 1}{\rho_{\rm p}}^{\rm s}+(2a_{\rm 1} +4a_{\rm 3})({\rho_{\rm p}}^{\rm s} + {\rho_{\rm n}}^{\rm s})]
{ m_{\rm s} \over 2f^{\rm 2}}  = 0  \]
Once the effective mass for nucleon is obtained from self consistent solution,
the total pressure density can be calculated as, \\
\begin{eqnarray}
p &=& {\gamma \over 6{\pi}^{\rm 2}} \sum_{\rm B}{\int_{\rm 0}}^{\rm k} 
dk k^{\rm 2} {k^{\rm 2} \over \sqrt{k^{\rm 2} + {{m_{\rm B}}^{\rm *}}^{\rm 2}}} - {1 \over
2}{m_{\rm \sigma}}^{\rm 2}{\sigma}^{\rm 2} \nonumber \\
&+& {1 \over 2}{m_{\rm \omega}}^{\rm 2}{\omega}^{\rm 2} + {1 \over 2}{m_{\rm 
\rho}}^{\rm 2}{\rho}^{\rm 2} - {1 \over 3}b{\sigma}^{\rm 3} - {1 \over 4}
c{\sigma}^{\rm 4} \nonumber \\
&+& {1 \over 3{\pi}^{\rm 2}} \sum_{\rm {leptons}}{\int_{\rm 0}}^{\rm k}
dk k^{\rm 2} {k^{\rm 2} \over \sqrt{k^{\rm 2} + {m_{\rm L}}^{\rm 2}}} \nonumber
\end{eqnarray}
The contribution to pressure from neutrino is given by, \\
\[ p_{\rm \nu} = {1 \over 24{\pi}^{\rm 2}}{{{\mu}_{\rm \nu}}_{\rm e}}^{\rm 4}
\]
\vskip 1.0cm
{\bf Results and Discussions} \\
\vskip\baselineskip
 We have studied the possibility of kaon condensation in nucleon
only matter (without hyperon) under the trapped neutrino condition. {\bf{ Figure
 Ia}} shows the relative particle fraction $\rho/\rho_{\rm B}$ in the case 
when neutrino is absent. It shows the proton-electron degenerated curve 
to be split into two because of muon production. When the electron fermi 
momentum reaches the muon mass, muons are being produced. In the absence of neutrino, the threshold density for muon production occurs just near the normal 
nuclear density $\simeq$ 1.05$\rho_{\rm 0}$. But the presence of 
neutrino shifts this
threshold density to much higher density. {\bf{Figure Ib}} shows that 
this threshold is shifted to $\rho$ $\simeq$ 5$\rho_{\rm 0}$ when neutrinos are being trapped 
in the matter for a typical value of final collapse $Y_{\rm {le}}$ = 0.3. \\
\parindent 3pt The electron neutrinos being produced in this system are pauli blocked, so more
neutrons will decay in order to preserve $Y_{\rm {le}}$ at constant value.
As a result, the matter becomes a protoneutron star. The presence of neutrino
shifts the threshold density for kaon condensation to much higher density   
$\rho$ $\simeq$  8$\rho_{\rm 0}$ for $Y_{\rm {le}}$ = 0.3 where normal threshold density
for possible kaon condensation is 4$\rho_{\rm 0}$ without neutrino trapping. \\
\parindent 3pt Here the kaon condensation process depends on two factors, 
the first being the behaviour of the scalar field $g_{\rm \sigma}{\sigma}$
in the matter and the second being the strangeness content of the nucleon 
related to the magnitude of $a_{\rm 3}m_{\rm s}$. 
The larger the magnitude of $a_{\rm 3}m_{\rm s}$, the lower is the value of 
threshold density $\rho_{\rm c}$. \\
\parindent 3pt  
 It is also seen that though neutrino plays a major role in determining 
threshold density for kaon condensation but the 
condensate amplitude grows rapidly and at that stage the leptons play only a
minor role only to preserve the $Y_{\rm {le}}$ constant.  
 Since the threshold density for possible kaon condensate is very high
, so this may not be present in the core of protoneutron star system. 
We have plotted the nucleon effective mass,  scalar field and electron chemical potential against
density  in {\bf{Figs. IIa \& IIb}} showing the behaviour in dense matter. 
The scalar field is less sensitive to density and almost constant in 
trapped neutrino matter which effectively decreases the kaon chemical potential.
For typical final collapse value of $Y_{\rm {le}}$ = 0.3, $\mu$ $\simeq$ 49 MeV
in contrast to $\mu$ $\simeq$ 108 MeV untrapped value. \\
\parindent 3pt
The electron capture processes which proceed with density becomes stopped
when trapped neutrinos settle into the matter. The immediate consequence is the decrease in $\mu$ values which significantly delay the kaon condensation until
high density. This feature is significant for the evolution of neutron star
at the early stage. Though kaon condensation softens the equation of state,
but in trapped case the delayed kaon condensation gives rise to more pressure.
As a result, the equation of state becomes more stiff in ths case. 
{\bf{ Fig. III}} compares the pressure with and without kaon condensation in
trapped neutrino matter with that pressure in case of free neutrino matter.
As long as neutrinos remain trapped in the matter, the overall pressure is very
high and gets softened when neutrinos diffuse out from the system. The $Y_{\rm {le}}$ drops out and threshold density for kaon condensation decreases. So kaon
could appear in the matter which in turn softens the EOS of the matter. This
may be the reason for delayed exploision mechanism of supernovae. \\
\vskip 1.0cm
{\bf{ Acknowledgement:}} \\
\parindent 3pt
The author is thankful to C.S.I.R for financial support during the course
of the work.

{\bf{ References:}
}
\vskip 0.2cm
\noindent 1.
 E.H.Gudmudsson and J.R. Buchler; Astrophys. J. {\bf 238} (1980)
717. \\
\noindent 2. N.K. Glendenning ; Astrophysical J. {\bf 293} (1985) 470. \\ 
(1996) 936. \\
\noindent 3. D.B. Kaplan and A.E. Nelson; Phys. Lett {\bf B 192} (1986)
409. \\
\noindent 3. V.Thorsson, M.Prakash and J. Lattimer; Nucl. Phys. {\bf A 572}
 (1994) 693. \\
\noindent 4. G.E. Brown, K. Kubodera, M. Rho and V. Thorsson; Phys. Lett.
 {\bf B 291} (1992) 355. \\
 11. \\
\noindent 5. V.Thorsson, M.Prakash and J. Lattimer; Nucl. Phys. {\bf A 572}
 (1994) 693. \\
\noindent 6. M. Prakash, T.L. Einswarth and J.M. Lattimer; Phys. Rev. Lett.
{\bf 61} (1988) 2518. \\
\noindent 7. P. Ellis, R. Knorren and M. Prakash; Phys. Lett. {\bf B 349}
 (1995) 11. \\
\noindent 8. M.Chiapparini, H. Rodrigues and S.B.Duarte; Phys. Rev.{\bf C54}
(1996) 936. \\
\vskip 0.5cm
\centerline{\bf{Figure Captions:}}
\vskip 0.2cm
\begin{enumerate}
\item {\bf{ Fig. Ia}}. The relative particle population of nucleons
as function of total
baryon density [ $\rho_{\rm B}$/$\rho_{\rm 0}$ , $\rho_{\rm 0}$ is equilibrium
nuclear density ] in neutrino free dense  matter ( without hyperon ) 
  
\item {\bf{ Fig. Ib}}. The relative particle population of nucleonsas function
of total 
baryon density for the value of electron-lepton fraction ${Y_{\rm l }}_{\rm e}$
under trapped neutrino  condition. 
\item {\bf { Fig. IIa}}. The nucleon effective masses, the kaon chemical
potential $\mu$ = $\mu_{\rm e}$ and the scalar field $g_{\rm \sigma}\sigma$
in free neutrino matter.
\item {\bf {Fig. IIb}}. The nucleon effective mass, the kaon chemical potential
$\mu$ = $\mu_{\rm e}$ - ${\mu_{\rm \nu}}_{\rm e}$ and the scalar field for
$Y_{\rm {le}}$ = 0.3
\item {\bf {Fig. III}}. Pressure with and without kaon condensation in free 
neutrino case compared with the pressure with kaon condensation under trapped
neutrino.
\end{enumerate}  
\vskip 1.0cm
\begin{center}  
{\bf {Table I. Coupling Constants used in this calculation}} \\
\begin{tabular}{|c|c|c|c|c|} 
\hline \hline
$(g_{\rm {\sigma}}/m_{\rm {\sigma}})^{\rm 2}$ & $(g_{\rm {\omega}}/m_{\rm 
{\omega}})^{\rm 2}$ & $(g_{\rm {\rho}}/m_{\rm {\rho}})^{\rm 2}$ & b & c \\ 
\hline \hline
10.138 & 13.285 & 4.975 & 0.003478 & 0.01328 \\ \hline \hline
\end{tabular}
\end{center}
\newpage
\begin{center}
\hbox{
\psfig{figure=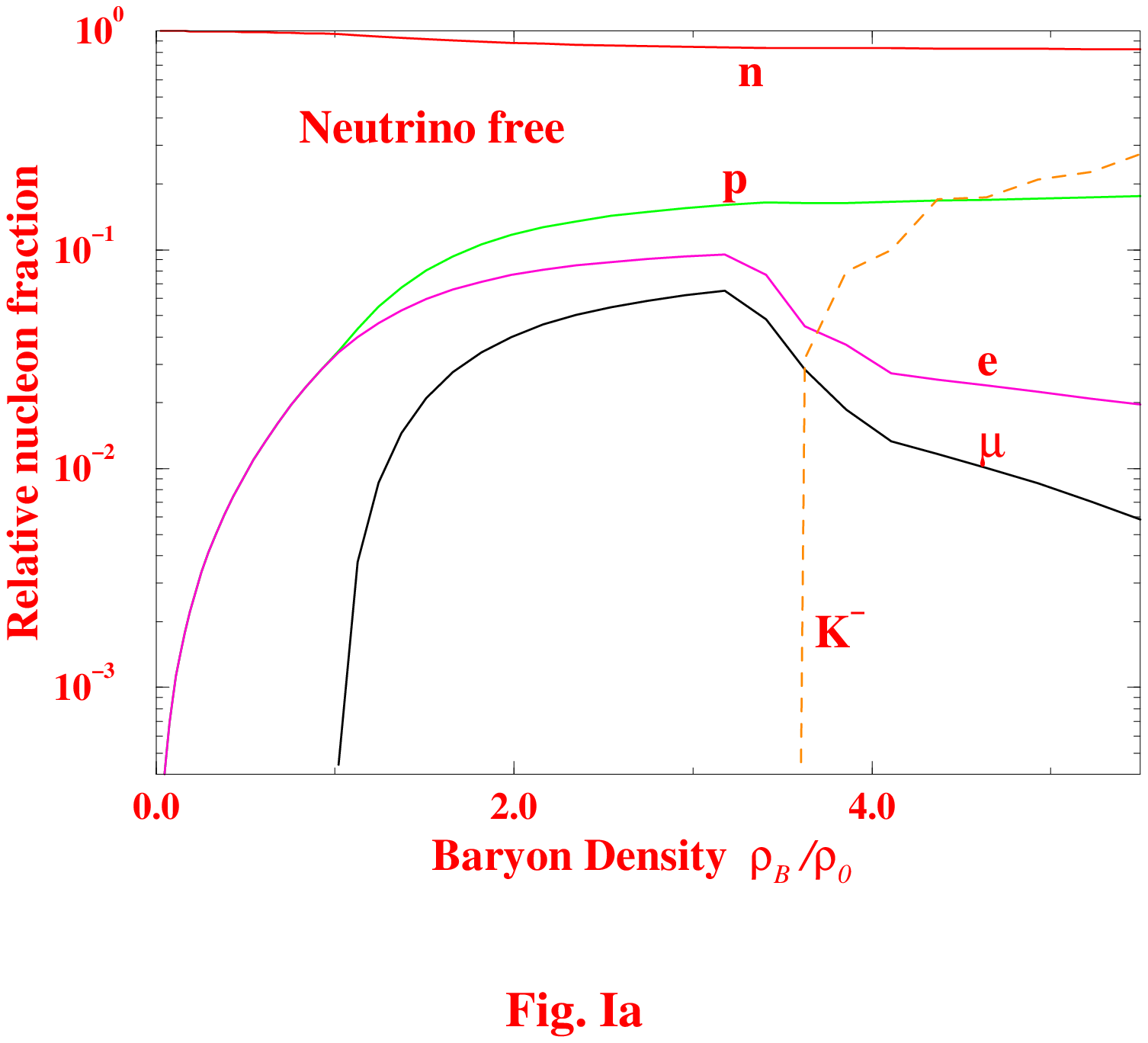}}
\newpage
\hbox{
\psfig{figure=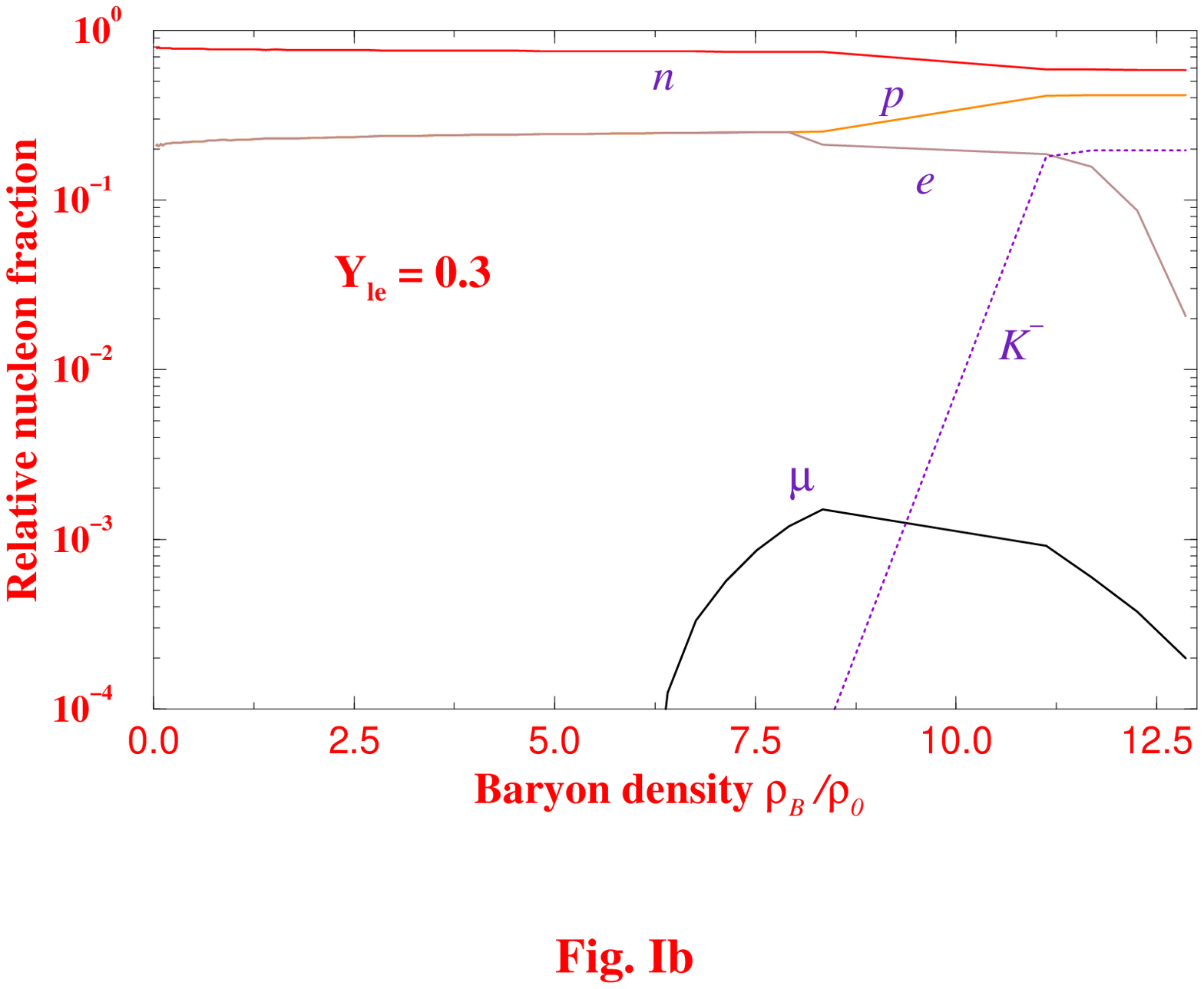}}
\newpage
\hbox{
\psfig{figure=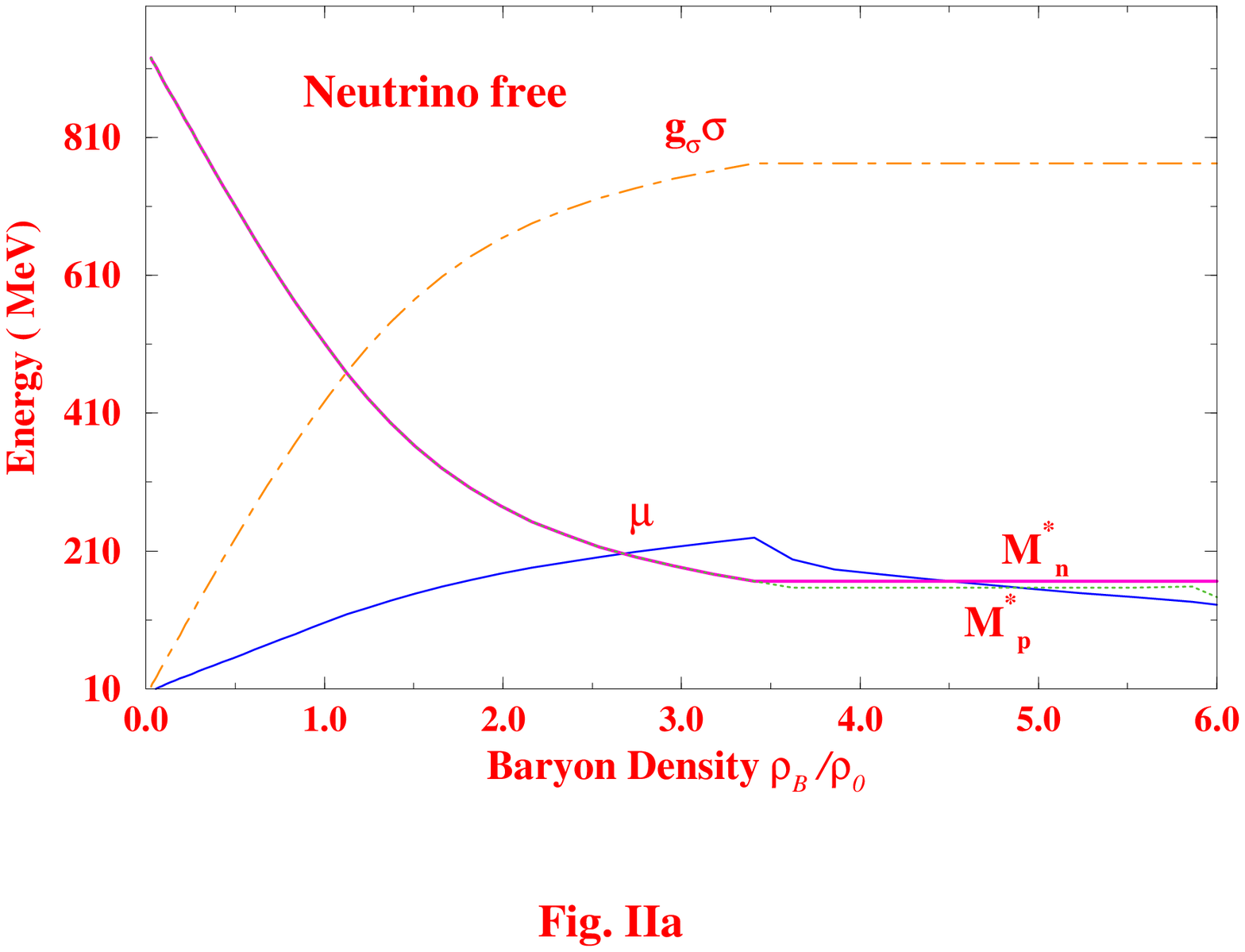}}
\newpage
\hbox{
\psfig{figure=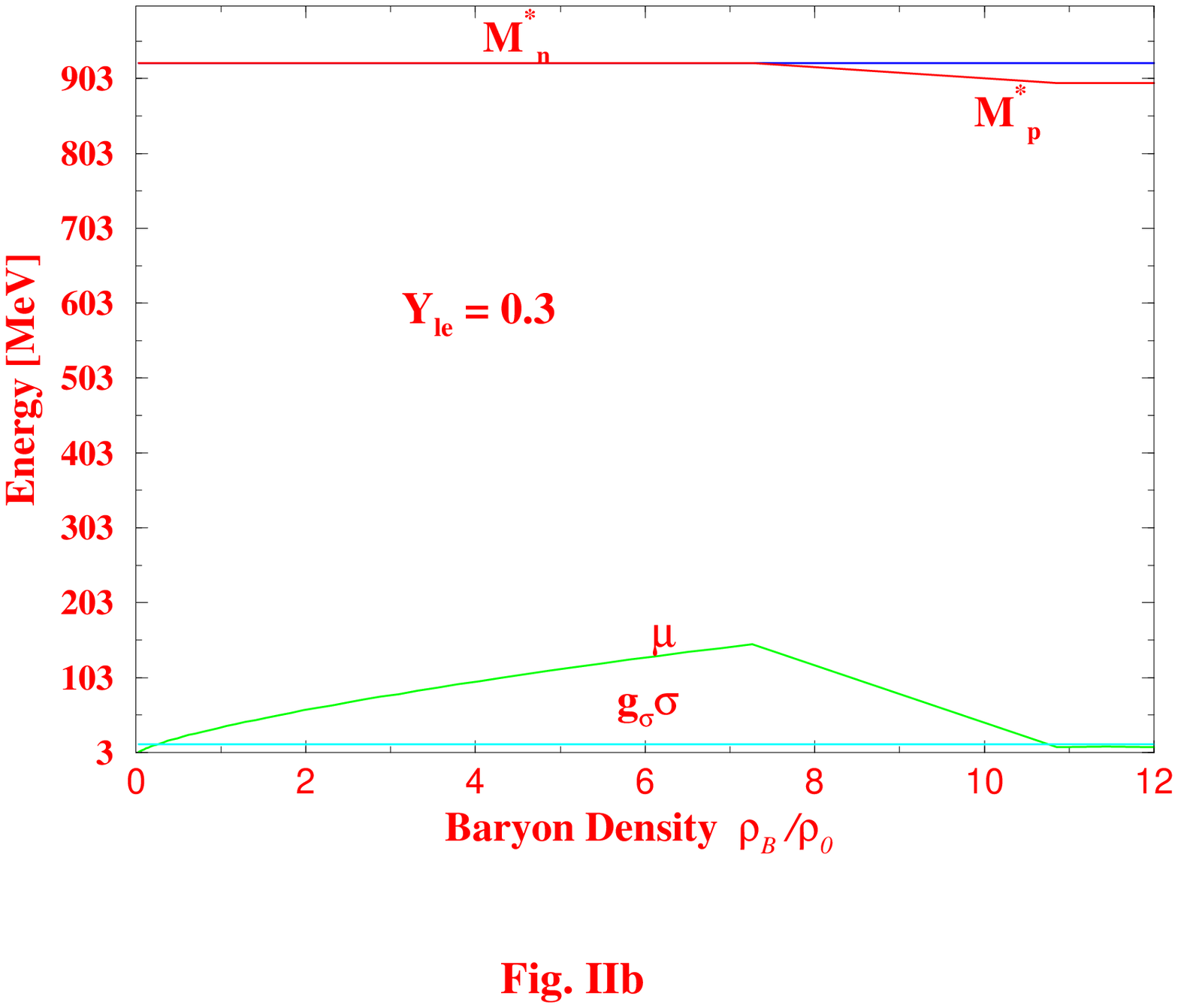}}
\newpage
\hbox{
\psfig{figure=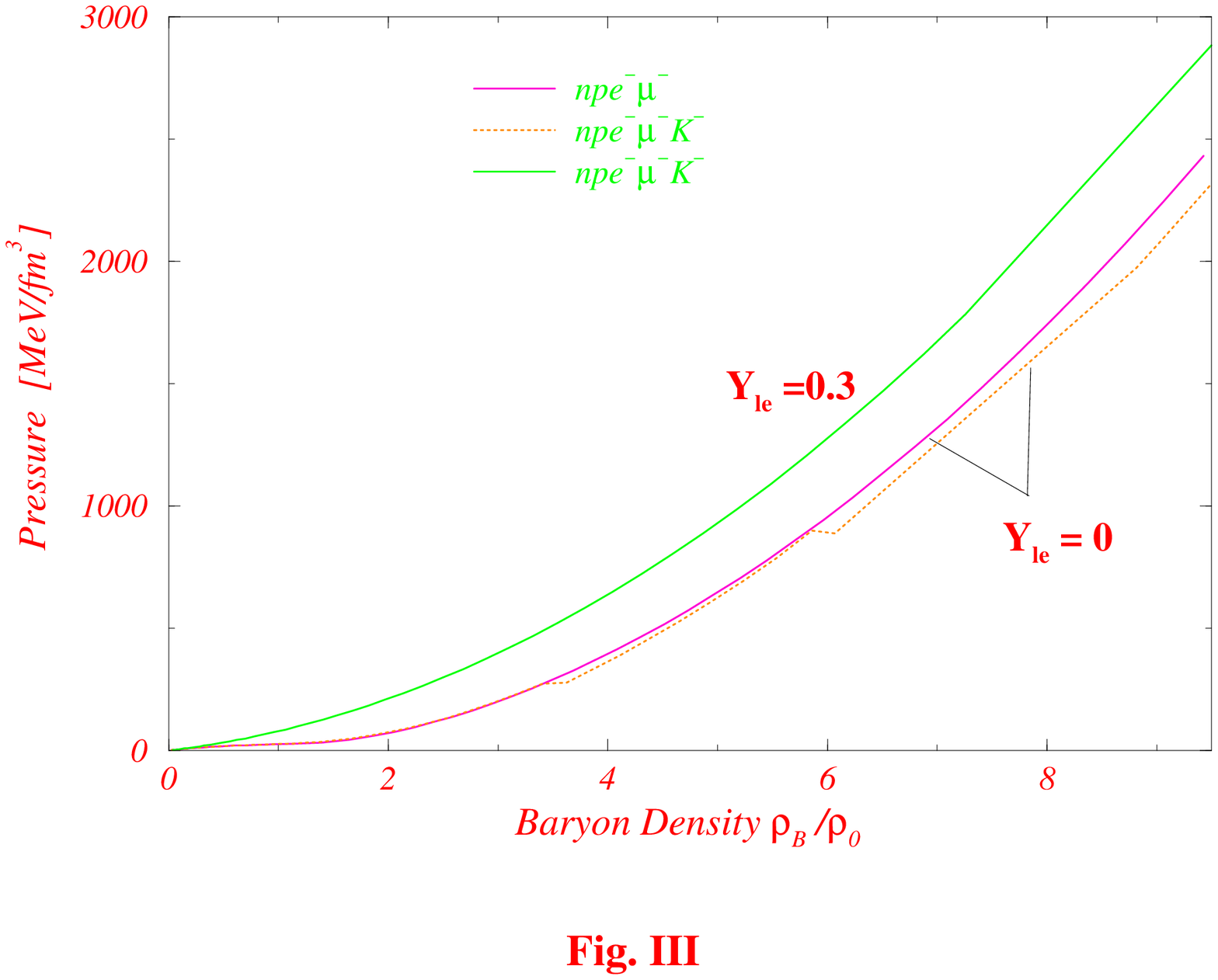}}
\end{center}

\end{document}